\documentclass[aps,prl,superscriptaddress,twocolumn,showpacs,floatfix]{revtex4}
\usepackage{epsfig}
\begin{document}
\title{Statistical Mechanics of Community Detection}

\author{J\"org Reichardt} 
\author{Stefan Bornholdt}
\affiliation{Institute for Theoretical Physics, University of Bremen, Otto-Hahn-Allee, D-28359 Bremen, Germany}
\date{\today}
\begin{abstract}
\noindent Starting from a general \textit{ansatz}, we show how community detection can be interpreted as finding the ground state of an infinite range spin glass. Our approach applies to weighted and directed networks alike. It contains the \textit{at hoc} introduced quality function from \cite{ReichardtPRL} and the modularity $Q$ as defined by Newman and Girvan \cite{Girvan03} as special cases. The community structure of the network is interpreted as the spin configuration that minimizes the energy of the spin glass with the spin states being the community indices. We elucidate the properties of the ground state configuration to give a concise definition of communities as cohesive subgroups in networks that is adaptive to the specific class of network under study. Further we show, how hierarchies and overlap in the community structure can be detected. Computationally effective local update rules for optimization procedures to find the ground state are given. We show how the \textit{ansatz} may be used to discover the community around a given node without detecting all communities in the full network and we give benchmarks for the performance of this extension. Finally, we give expectation values for the modularity of random graphs, which can be used in the assessment of statistical significance of community structure. 
\end{abstract}
\maketitle

\section{Introduction}
\noindent The amount of empirical information that scientists from all disciplines are dealing with is constantly increasing.  And so is the need for robust, scalable, and easy to use clustering techniques for data abstraction, dimensionality reduction, or visualization for many scientists performing exploratory data analysis \cite{Everitt, JainReview}. A basic objective is to group objects which are similar together, and dissimilar objects apart. But already the question of how to measure similarity/dissimilarity is a subject of discussion \cite{JainReview}. Two main approaches to clustering are identified in the literature \cite{JainReview}: On one hand there is hierarchical clustering where the data set is grouped into a hierarchy of clusters from single items to the whole data set. Data points are either joined successively in an agglomerative manner starting from the closest pair of data points or the data set is recursively partitioned into two parts, an approach which is called divisive. On the other hand, there is partitional clustering, where the data set is directly partitioned into $k$ different clusters usually optimizing some quality function. The number of clusters $k$ is either an input parameter of the algorithm or found by the clustering procedure itself. Transforming the similarity matrix into a graph by \textit{e.g.}~thresholding, the clustering problem can be tackled from a graph partitioning point of view. These approaches apply directly to networks or relational data sets where the proximity information is given as a set of pairwise relations, \textit{i.e.} the edges of the network. The problem is then approached by a min-cut technique that partitions a connected graph into two parts minimizing the number of edges to cut \cite{KernighanLin, FiducciaMattheyses,DingMinMax}. These approaches, however, suffer greatly from being very skewed as the min-cut is usually found by cutting off only a very small subgraph \cite{Cheng1991}. A number of penalty functions have been suggested to overcome this problem and balance the size of subgraphs resulting from a cut. Among these are ratio cuts \cite{Cheng1991, Hagen1992}, normalized cuts \cite{Shi2000} or min-max cuts \cite{DingMinMax}.     

Though today the development of these methods lies mainly in the realm of computer science, the relations between information theory and statistical physics \cite{Jaynes1, Jaynes2} have brought about a number of such methods that are based on principles from statistical mechanics or analogies with physical models.  When using spin models for clustering of multivariate data, the similarity measures are translated into coupling strengths and either dynamical properties such as spin-spin correlations are measured or energies are interpreted as quality functions. A ferromagnetic Potts model has been  applied successfully by Blatt \textit{et al.\ } \cite{Blatt}. Bengtsson \cite{Bengtsson} has used an anti-ferromagnetic Potts model with the number of clusters $k$ as input parameter and the assignment of spins in the ground state of the system defines the clustering solution.  

In recent years, renewed interest in the graph clustering problem from the physics community has come under the term of  ``community detection''.  As communities, one generally understands subsets of nodes that are more densely interconnected among each other than with the rest of the network. Sparked by the work of Girvan and Newman \cite{Girvan}, a number of other authors have developed new algorithms for this problem that take very different approaches. The recent reviews by Newman \cite{Newman} and Danon \textit{et al.}~ \cite{GuileraReview} may serve as introductory reading and include methodological overviews and comparative studies of the performance of different algorithms, including the one presented by the authors in \cite{ReichardtPRL}. In this article, we intend to set the basis for a unified framework under which community detection may be viewed and which helps in understanding the underlying properties of the problem. 

First, we will show that the problem of community detection can be mapped onto finding the ground state of an infinite ranged Potts spin glass via a simple first principles \textit{ansatz} by combining the information from both present and missing links. The energy of the spin system is equivalent to the quality function of the clustering with the spins states being the group indices. In the above taxonomy of clustering procedures, this corresponds to a partitional method with the number of clusters determined automatically by the algorithm as the number of occupied spin states. A single parameter  $\gamma$ relates the weight given to missing and existing links in the quality function and allows for an assessment of overlapping and hierarchical community structures. Thereby, we can bridge the gap between hierarchical and partitional clustering and conclude to which extent the cluster structure of the network is hierarchical or not.  

In contrast to methods based on dynamical properties of the spin system that measure correlations between spins, such as the super-paramagnetic (SPC) Potts clustering introduced by Blatt \textit{et. al.,\ } mapping the problem to a ground state bears several advantages. First, it is computationally less demanding, because we do not have to keep track of an $N\times N$ correlation matrix of spin states. Rather, every spin only carries its most probable community index. If a probabilistic extension of the method is required, an analysis of the overlap of the community structures in different local minima of the Hamiltonian can be performed as done in \cite{ReichardtPRL}. Second, the properties of the ground state spin configuration lead to a direct interpretation of the result in terms of graph theoretical measures, which give an exact definition of what a ``community'' is in this framework. The interpretation of the parameter $\gamma$ in the evaluation of hierarchy and overlap is much clearer than the interpretation of the temperature in SPC. Third, the zero temperature energy can be calculated analytically which allows to give expectation values of the modularity and assess the clustering tendency of the graph under study.   

For a natural choice of parameters, we recover the ``modularity'' defined by Girvan and Newman \cite{Girvan03} from our \textit{ansatz} as well as the \textit{ad hoc} introduced quality function from \cite{ReichardtPRL}. Then we will derive a number of graph structural properties that define the term ``community'' from the fact that valid community structures correspond to minima of the energy landscape of the system. We compare this definition to other possibilities from the literature. We then show, how hierarchical and overlapping community structures can be discovered in this framework. Even though the quality function resembles an infinite ranged spin glass with couplings between all pairs of nodes, we show how efficient minimization routines can be implemented that only need to consider interactions along the links in the network and some global book keeping. This makes the use of the method feasible even for large systems. Furthermore, we show how a method of finding the community around a given node can be developed in this general framework and give benchmarks for this method. All clustering procedures will find clusters even when applied to random data. Hence, in the last part of the paper, we focus on the statistical significance of community detection. We show, how community detection is related to graph partitioning and that when community detection is applied to random graphs, equally sized communities are found. From the known results for the cut size of graph partitionings we can calculate expectation values for the modularity of random graphs which have to be exceeded by any data set that is to be called truly modular.   

\section{Derivation of the Hamiltonian}
\noindent For the term ``community'' or ``cluster'' or ``cohesive subgroup'' a number of different and sometimes conflicting definitions exist \cite{GuileraReview}. All of them have in common that communities are understood as groups of densely interconnected nodes that are only sparsely connected with the rest of the network. Any quality function for an assignment of nodes into communities should therefore follow the simple principle: group together what is linked, keep apart what is not. From this, we find four requirements of such a quality function: it should a.) reward internal edges between nodes of the same group (in the same spin state) and b.) penalize missing edges (non-links) between nodes in the same group. Further, it should c.) penalize existing edges between different groups (nodes in different spin state) and d.) reward non-links between different groups. This leads to the following function: 
\begin{widetext}
\begin{eqnarray}
\mathcal{H}\left(\{\sigma\}\right) & = &-\sum_{i\neq j}a_{ij}\underbrace{A_{ij}\delta(\sigma_i,\sigma_j)}_{\mbox{internal links}}
+\sum_{i\neq j}b_{ij}\underbrace{(1-A_{ij})\delta(\sigma_i,\sigma_j)}_{\mbox{internal non-links}}\\
& & +\sum_{i\neq j}c_{ij}\underbrace{A_{ij}(1-\delta(\sigma_i, \sigma_j))}_{\mbox{external links}}
-\sum_{i\neq j}d_{ij}\underbrace{(1-A_{ij})(1-\delta(\sigma_i, \sigma_j))}_{\mbox{external non-links}}\nonumber
\label{origHamiltonian}
\end{eqnarray}
\end{widetext}

in which $A_{ij}$ denotes the adjacency matrix of the graph with $A_{ij}=1$, if an edge is present and zero otherwise, $\sigma_i\in\{1,2,...,q\}$ denotes the spin state (or group index) of node $i$ in the graph and $a_{ij},b_{ij},c_{ij},d_{ij}$ denote the weights of the individual contributions, respectively. The number of spin states $q$ determines the maximum number of groups allowed and can, in principle, be as large as $N$, the number of nodes in the network. Note, that not all group indices have to be used necessarily in the optimal assignment of nodes into communities, as some spin states may remain unpopulated in the ground state. If links and non-links are each weighted equally, regardless whether they are external or internal, \textit{i.e.}~$a_{ij}=c_{ij}$ and $b_{ij}=d_{ij}$, then it is enough to consider the internal links and non-links. It remains to find a sensible choice of weights $a_{ij}$ and $b_{ij}$, preferably such that the contribution of links and non-links can be adjusted through a parameter. As we will see, a convenient choice is $a_{ij}=1-\gamma p_{ij}$ and $b_{ij}=\gamma p_{ij}$, where $p_{ij}$ denotes the probability that a link exists between node $i$ and $j$, normalized, such that $\sum_{i\neq j}p_{ij}=2M$. For $\gamma=1$ this leads to the natural situation that the total amount of energy that can possibly be contributed by links and non-links is equal: $\sum_{i\neq j}A_{ij}a_{ij}=\sum_{i\neq j}(1-A_{ij})b_{ij}$. For weighted networks this approach is generalized in a straight-forward manner by using a weighted adjacency matrix $W_{ij}$. In case of a directed network with a non-symmetric adjacency matrix $A_{ij}\neq A_{ij}$, one can construct a symmetric representation of network introducing $\tilde{A}_{ij}=1/2(A_{ij}+A_{ji})$ and $\tilde{p}_{ij}=1/2(p_{ij}+p_{ji})$. In this article, we will only deal with undirected, unweighted adjacency matrices.    
Our choice of the weights allows us to further simplify the Hamiltonian (\ref{origHamiltonian}): 
\begin{equation}
\mathcal{H}\left(\{\sigma\}\right)=-\sum_{i\neq j}\left(A_{ij}-\gamma p_{ij}\right)\delta(\sigma_i,\sigma_j).
\label{EasyHam}
\end{equation}
This represents a spin glass with couplings $J_{ij}=A_{ij}-p_{ij}$ between all pairs of nodes: ferromagnetic where links between nodes exist and anti-ferromagnetic where links are absent. 

Depending on the graph under study, one can assume different expressions for $p_{ij}$.  The Hamiltonian (\ref{EasyHam}) is effectively comparing the true distribution of links in the graph under study with the expected distribution given by a particular null model which defines $p_{ij}$. With this in mind, we can rewrite (\ref{EasyHam}) in the following two ways:
\begin{equation}
\mathcal{H}\left(\{\sigma\}\right)=-\sum_s\left(m_{ss}-\gamma [m_{ss}]_{p_{ij}}\right)=-\sum_s c_{ss}
\label{InternalForm}
\end{equation}
and
\begin{equation}
\mathcal{H}\left(\{\sigma\}\right)=\sum_{s<r}\left(m_{rs}-\gamma [m_{rs}]_{p_{ij}}\right)=\sum_{s<r}a_{rs}.
\label{ExternalForm}
\end{equation}
Here, the sum runs over the $q$ spin states and $m_{rs}$ denotes the number of edges between spins in group $r$ and $s$. Consequently, the number of internal edges of group $s$ is denoted by $m_{ss}$. The symbol $[\cdot]_{p_{ij}}$ denotes an expectation value under the assumption of a link distribution $p_{ij}$, given the current assignment of spins. 

In equation (\ref{InternalForm}) and (\ref{ExternalForm}) we have also introduced the coefficients of ``cohesion'' $c_{ss}$ and ``adhesion'' $a_{rs}$ to our network terminology, which measure the difference between realized and expected internal links or realized and expected external links, respectively.  Note, that both depend on the choice of the model of connectivity $p_{ij}$ and the parameter $\gamma$. The choice of a particular form of $p_{ij}$ allows for the adaptation of the quality function to the specific problem under study and hence allows for the comparison of the quality function for graphs with different topology. The only restriction on $p_{ij}$ is that the number of expected edges between and within groups is an extensive quantity, \textit{i.e.}~$[m_{13}]_{p_{ij}}+[m_{23}]_{p_{ij}}=[m_{1+2,3}]_{p_{ij}}$ for all choices of disjoint groups $n_1,n_2$ and $n_3$ and $[m_{33}]_{p_{ij}}=[m_{11}]_{p_{ij}}+[m_{22}]_{p_{ij}}+[m_{12}]_{p_{ij}}$ for all groups $3$ with proper subgroups $n_1$ and $n_2$ of empty intersection and union $n_3$.  Using these equalities, we can give a relation for the coefficient of cohesion of a group of nodes $n_s$ and two proper subsets $n_{s1}$ and $n_{s2}$ with empty intersection and union $n_s$. It is easy to prove, that
\begin{equation}
c_{ss}=c_{11}+c_{22}+a_{12},
\end{equation}
where $c_{11}$ and $c_{22}$ are the coefficients of cohesion of the respective subsets $n_{s1}$ and $n_{s2}$, and $a_{12}$ is the coefficient of adhesion between $n_{s1}$ and $n_{s2}$. Equivalently, we can write for the adhesion coefficients with $n_2$ of two groups $n_{r1}$ and $n_{r2}$ with union $n_r$ and empty intersection
\begin{equation}
a_{rs}=a_{1s}+a_{2s}.
\end{equation}
 
Two exemplary choices of link distribution models $p_{ij}$ shall illustrate the above. The simplest choice is to assume every link equally probable with probability $p_{ij}=p$ which leads naturally to 
\begin{equation}
[m_{ss}]_p=p \frac{n_s(n_s-1)}{2}\hspace{0.5cm}\mbox{ and }\hspace{0.5cm}[m_{rs}]_p=p n_rn_s,
\label{MSSMRS1}
\end{equation}
with $n_r$ and $n_s$ denoting the number of spins in state $r$ and $s$, respectively. This choice of model leads to the Hamiltonian originally quoted in Ref. \cite{ReichardtPRL}:
\begin{equation}
\mathcal{H}\left(\{\sigma\}\right)=-\sum_{i,j\in E}\delta(\sigma_i,\sigma_j)+\gamma p\sum_s^q\frac{n_s(n_s-1)}{2}.
\label{OrigHam}
\end{equation}
Here, the first sum runs over all edges and only internal edges contribute. Equivalently, we can write (\ref{OrigHam}) in terms of external edges:
\begin{equation}
\mathcal{H}(\{\sigma\})=\sum_{i,j\in E}(1-\delta(\sigma_i,\sigma_j))-\gamma p\sum_{r<s}^q n_rn_s,
\label{OrigHam2}
\end{equation}
where only edges between different groups contribute to the first sum. 
We see that both, (\ref{OrigHam}) and (\ref{OrigHam2}), compare the actual value of internal or external edges with its respective expectation value under the assumption of equally probable links and given community sizes.  

A second choice for $p_{ij}$ may take into account, that the network does exhibit a particular degree distribution. Since links are in principle more probable between nodes of high degree, links between these nodes get a lower weight. We may write:
\begin{equation}
p_{ij}=\frac{k_ik_j}{2M},
\end{equation} 
which takes this fact and the degree distribution into account. Note that it is possible to also include degree-degree correlations or any other form of prior knowledge about $p_{ij}$ at this point. With these expressions we have: 
\begin{equation}
[m_{ss}]_{p_{ij}}=\frac{1}{2M}\frac{K_s^2}{2}\mbox{ and }[m_{rs}]_{p_{ij}}=\frac{1}{2M}K_rK_s.
\label{MSSMRS2}
\end{equation}
Here, $K_s$ is the sum of degrees of nodes in spin state $s$ and plays the role of the occupation numbers in equation (\ref{OrigHam}). Using these expressions, we can also write the Hamiltonian (\ref{EasyHam}) in a form similar to (\ref{OrigHam}):
\begin{equation}
\mathcal{H}\left(\{\sigma\}\right)=-\sum_{i,j\in E}\delta(\sigma_i,\sigma_j)+\frac{\gamma}{2M}\sum_{s}^q\frac{K_s^2}{2}.
\label{AltHam}
\end{equation}
 Again, we give an equivalent formulation in terms of external rather than internal edges similar to (\ref{OrigHam2}):
\begin{equation}
\mathcal{H}\left(\{\sigma\}\right)=\sum_{i,j\in E}(1-\delta(\sigma_i,\sigma_j))-\frac{\gamma}{2M}\sum_{r< s}^qK_rK_s.
\label{AltHam2}
\end{equation}
For $\gamma=1$ and the model $p_{ij}=k_ik_j/2M$, we can derive
\begin{equation}
2c_{ss}+\sum_{r,r\neq s}a_{sr}=0.
\label{SpecialCase}
\end{equation}
Furthermore, the cohesion is negative ($c_{ss}<0$) if $n_s$ consists of only one single node. We see, that there must always exist a group of nodes $n_r$, to which this node has positive adhesion.  Groups of only one node do not exist. We stress that relation (\ref{SpecialCase}) and the conclusions just drawn do not hold for $\gamma\neq 1$ or $p_{ij}=p$.

We see, that even though we are dealing with an infinite range spin glass with couplings between all pairs of nodes, one only needs to consider the ferromagnetic interactions along the links and the occupation numbers or the sum of node degrees of the individual spin states. This makes it easy to implement an efficient minimization routine for this Hamiltonian. It should be noted, that both the formulations (\ref{OrigHam}), (\ref{OrigHam2}) and (\ref{AltHam}), (\ref{AltHam2}) are equivalent in case of a network with fixed connectivity. 

\section{Equivalence with Newman-Girvan Modularity}
\noindent Comparing the performance in retrieving a known community structure from computer generated test networks may be used as a benchmark for different community detection algorithms. Alternatively, many authors have given values of the quality function $Q$ defined by Newman and Girvan as ``modularity'' \cite{Girvan03} as a global, comparative, objective measure of how good a community structure found by an algorithm is. Alternative formulations focussing on the local aspects of community structure also exist, such as that of ``local modularity'' introduced by Muff \textit{et al.}~\cite{Muff}. Newman and Girvans's modularity measure can be written as \cite{Girvan03}:
\begin{equation}
Q = \sum_s e_{ss}-a_s^2\mbox{,\hspace{0.5cm}  with }a_s = \sum_r e_{rs}.
\end{equation}
Here, $e_{rs}$ is the fraction of links that fall between nodes in group $r$ and $s$, \textit{i.e.}~the probability that a randomly drawn link connects a node in group $r$ to one in group $s$. The probability that a link has one end in group $s$ is expressed by $a_s$. From this, we expect a fraction of $a_s^2$ links to connect nodes in group $s$ among themselves. Newman's modularity measure hence compares the actual link density in a community with an expectation value. One can write this modularity in a slightly different way following \cite{NewmanLarge}:
\begin{equation}
Q  =  \frac{1}{2M}\sum_{i\neq j}\left(A_{ij}-\frac{k_ik_j}{2M}\right)\delta(\sigma_i,\sigma_j).\\
\label{ModularityHam}
\end{equation}
This already resembles (\ref{EasyHam}) when $p_{ij}$ takes the form $k_ik_j/2M$. It is now clear, that we can write: 
\begin{equation}
Q=-\frac{1}{M}\mathcal{H}(\{\sigma\})
\end{equation}
with the Hamiltonian (\ref{EasyHam}) and $\gamma=1$. Therefore, maximum modularity is reached, when the Hamiltonian (\ref{EasyHam}) with $p_{ij}=k_ik_j/2M$ or equivalently (\ref{AltHam}) or (\ref{AltHam2}) with $\gamma=1$ are minimal. To maximize the modularity of a community structure is hence equivalent to finding the spin configuration that minimizes these Hamiltonians. This form of writing the modularity $Q$ is much simpler than the one given by Guimera \textit{et al.}~\cite{Guimera}, which also involves 3- and 4-spin interactions. We will see below, that using this form, we can give efficient update rules that allow the direct optimization of the modularity even on very large networks.   

\section{Properties of the Hamiltonian and its Ground State}
\noindent Having mapped the problem of community finding onto finding the ground state configuration of a spin glass, we can investigate the properties of this minimum energy spin configuration. These properties will provide us with a \textit{definition} of what a community is in the framework of maximizing a quality function. These properties will apply to any local minimum of the Hamiltonian as well, such that we can interpret these local minima as alternative community structures. Inspection of the total energy landscape and comparison of global and local minima and the respective community structure will then provide insight into the clustering tendency of the network. Obviously, the more local minima with little overlap but energies comparable to the global minimum there are, the more spin glass like the energy landscape is, the less the network shows a truly modular structure.

Since the Hamiltonians are all additive with respect to the different communities, \textit{i.e.}~the numbers of edges and the corresponding expectation values are extensive, they can be seen as independent entities and we can treat a single community independently from the rest of the network. The configuration space over which the Hamiltonian is minimized is a discrete space. Once we have defined a move set that is ergodic in this discrete space, a (local) minimum of the Hamiltonian (with respect to this move set) is defined as a configuration for which none of the steps from the move set leads to a lower energy. It is sufficient to consider only one move: change a group of nodes $n_1$ from spin state $s$ to spin state $r$. The change in energy for this move in configuration space is:
\begin{equation}
\Delta\mathcal{H}=a_{1,s\backslash 1}-a_{1r}.
\end{equation} 
Here $a_{1,s\backslash 1}$ is the adhesion of $n_1$ with its complement in $n_s$ and $a_{1r}$ is the adhesion of $n_1$ with $n_r$. If we move $n_1$ to a previously unpopulated spin state, then $\Delta\mathcal{H}=a_{1,s\backslash 1}$. This move corresponds to dividing group $n_s$. Furthermore, if $n_1=n_s$, we have $\Delta\mathcal{H}=-a_{sr}$, which corresponds to joining groups $n_s$ and $n_r$. For a spin configuration to be a local minimum of the Hamiltonian, there must not exist a move of this type that leads to a lower energy. It is clear that some moves may not change the energy and are hence called neutral moves. In case of equality $a_{1,s\backslash 1}=a_{1,r}$ and $n_r$ being a community itself, we say that communities $n_s$ and $n_r$ have an overlap of the nodes in $n_1$.   

For a community defined as a group of nodes with the same spin state in a spin configuration that makes the Hamiltonian minimal, we then have the following properties:
\begin{enumerate}
 \item Every proper subset $n_1$ of a community $n_s$ has a maximum coefficient of adhesion with its complement in the community compared to the coefficient of adhesion with any other community ($a_{1,s\backslash 1}=\mbox{max}$).  
\item The coefficient of cohesion is non-negative for all communities ($c_{ss}\geq 0$).
\item The coefficient of adhesion between any two communities is non-positive ($a_{rs}\leq 0$).  
\end{enumerate}

\noindent The first property is proven by contradiction from the fact that we are dealing with a spin configuration that makes the Hamiltonian minimal.  We also see immediately that every proper subset $n_1$ of a community $n_s$ must have a non-negative adhesion with its complement $n_{s\backslash 1}$ in the community.  In particular this is true for every single node $l$ in $n_s$ ($a_{l,s\backslash l}\geq 0$). Then we can write $\sum_{l\in n_s}a_{l,s\backslash l}\geq 0$. Since  $\sum_{l\in n_s}m_{l,s\backslash l}=2m_{ss}$ and  $\sum_{l\in n_s}[m_{l,s\backslash l}]_{p_{ij}}=2[m_{ss}]_{p_{ij}}$, this implies $c_{ss}\geq 0$ for all communities $s$ and proves the second property. The third property is proven by contradiction again.
Again, we stress that for $\gamma=1$ and $p_{ij}=k_ik_j/2M$, no community is formed of a single node due to condition (\ref{SpecialCase}).
The last two properties can be summarized in the following inequality which provides an intuition about the significance of the parameter $\gamma$: 
\begin{equation}
c_{ss}\geq 0\geq a_{rs}\hspace{0.5cm}\forall r,s.
\end{equation}
Assuming a constant link probability, we can rewrite this inequality in order to relate  the inner link density of a community and the outer link density between communities with an average link density:
\begin{equation}
\frac{2m_{ss}}{n_s(n_s-1)}\geq\gamma p\geq\frac{m_{rs}}{n_rn_s}\hspace{0.5cm}\forall r,s.
\end{equation} 
We see, that $\gamma p$ can be interpreted as a threshold between inner and outer link density under the assumption of a constant link probability. The above definition of what a community is adapts itself to any network, since the specific network model is encoded in the definition of cohesion and adhesion. This makes it possible to compare community structures of networks with different topology.

\section{Simple divisive and agglomerative approaches to modularity maximization}
\noindent Hierarchical clustering techniques can be dichotomized into divisive and agglomerative approaches \cite{JainReview}. We will show, how a simple recursive divisive approach and an agglomerative approach may be implemented and where they fail. 

In the present framework, a hierarchical divisive algorithm would mean to construct the ground state of the q-state Potts model by recursive partitioning the network into two parts according to the ground state of a 2-state Potts or Ising system. This procedure would be computationally simpler and result directly in a hierarchy of clusters due to repetition of the procedure on the parts until the total energy cannot be lowered anymore. Such a procedure would be justified, if the ground state of the q-state Potts Hamiltonian and the repeated application of the Ising system cut the network along the same edges. We will derive a condition under which this can be ensured. 

\begin{figure}
\includegraphics[width=8cm]{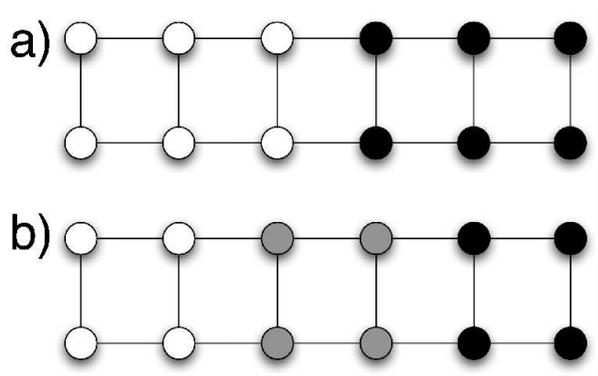}
\caption{Illustration of the problem of recursive bi-partitioning. The ground state of the Hamiltonian with only $2$ possible spin states, as shown in a.),  would cut through one of the communities that are found when allowing $3$ spin states as shown in b.).}
\label{BiPartProb}
\end{figure}

In order for this recursive approach to work, we must ensure that the ground state of the 2-state Hamiltonian never cuts though a community as defined by the q-state Hamiltonian. Assume a network made of three communities $n_1$, $n_2$ and $n_3$ as defined by the ground state of the q-state Hamiltonian. For the bi-partitioning, we now have two possible scenarios. Without loss of generality, the cut is made either between $n_2$ and $n_1+n_3$ or between $n_1$, $n_2$ and $n_3=n_a+n_b$, parting the network into $n_1+n_a$ and $n_2+n_b$. Since the former situation should be energetically lower for the algorithm to work, we arrive at the condition that 
\begin{equation}
m_{ab}-[m_{ab}]_{p_{ij}}+m_{1b}-[m_{1b}]_{p_{ij}}>m_{2b}-[m_{2b}]_{p_{ij}}, 
\label{BiPartCond}
\end{equation}
which must be valid for all subgroups $n_a$ and $n_b$ of community $n_3$. 
Since $n_3$ is a community, we further know, that $m_{ab}-[m_{ab}]_{p_{ij}}>m_{1b}-[m_{1b}]_{p_{ij}}$ and $m_{ab}-[m_{ab}]_{p_{ij}}>m_{2b}-[m_{2b}]_{p_{ij}}$.  Though $m_{ab}-[m_{ab}]_{p_{ij}}>0$, since $n_3$ is a community, $m_{1b}-[m_{1b}]_{p_{ij}}<0$ and $m_{2b}-[m_{2b}]_{p_{ij}}<0$ for the same reason and hence condition (\ref{BiPartCond}) is not generally satisfied. Figure \ref{BiPartProb} illustrates a counter example. Assuming $p_{ij}=p$, the link probability in the network. The upper part a.) of the figure shows the ground state of the system when using only two spin states. Part b.) of the figure shows the ground state of the system without constraints on the number of spin states, resulting in a configuration of $3$ communities. We see that the bi-partitioning approach would have cut through one of the communities in the network. Recursive bi-partitionings cannot generally lead to an optimal assigment of spins that maximizes the modularity.  

\begin{figure}
\includegraphics[width=7cm]{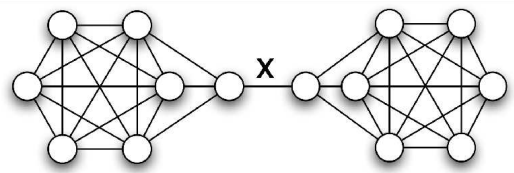}
\caption{Example network for which an agglomerative approach of grouping together nodes of maximal adhesion will fail. Starting from an assignment of different spin states to every node, the largest adhesion is found for the nodes connected by edge $x$ and the nodes connected by $x$ are grouped together first by the agglomerative procedure. However, it is clearly seen, that $x$ should lie between different groups.}
\label{SimpleMaxCounterEx}
\end{figure}

In \cite{FastGN} Newman has introduced a fast greedy strategy for modularity maximization. It effectively corresponds to a simple nearest neighbor agglomerative clustering of the network where the adhesion coefficient $a_{rs}$ is used as similarity measure between. Newman's algorithms initially assigns different spin states to every node and then proceeds by grouping those nodes together that have the highest coefficient of adhesion.  As Figure \ref{SimpleMaxCounterEx} shows, this approach fails, if the links between two communities connect nodes of low degree. The network consists of $14$ nodes and $37$ links. Is is clearly seen that in the ground state, the network consists of two communities and edge $x$ lies between them. However, when initially assigning different spin states to all nodes, the adhesion between the nodes connected by $x$ is largest: $a=1-16/2M$, since the product of degrees at this edge is lowest. Therefore, the agglomerative procedure described is misled into grouping together the nodes connected by $x$ already in the very first step. Furthermore, it is clear that in a network, where all nodes have the same degree initially, all edges connect nodes of the same coefficient of adhesion. In this case, it cannot be decided, which nodes to group together in the first step of the algorithm at all. It was shown by Newman, that the approach does deliver good results in benchmarks using computer generated test networks. The success of this approach depends of course on whether or not the misleading situations have a strong effect on the final outcome of the clustering. In the example shown, after grouping together the nodes at the end points of $x$, the algorithm will then proceed to further adding nodes from only one of the two communities linked by $x$. Hence, the initial mistake persists, but does not completely destroy the result of the clustering. 

\section{Comparison with other definitions of communities}
\noindent We have defined the term community as a set of nodes having properties $1.)$ through $3.)$. Compared with the many definitions of community in the sociological literature \cite{WassermanFaust}, this definition is most similar to that of an ``LS-set''. An LS-set is a set of nodes $S$ in a networks, such that each of its proper subsets has more links to its complement in $S$ than to the rest of the network \cite{Seidmann}. Previously, Radicchi \textit{et al.}~\cite{Radicchi} had given a definition of community  ``in a strong sense'' as a set of nodes $V$ with the condition $k_i^{in}>k_i^{out}, \forall i\in V$, \textit{i.e.}~every node in the group has more links to other members of the group than to the rest of the network. In the same manner, they define a community in a ``weak sense'', as a set of nodes $V$ for which $\sum_{i\in V}k_i^{in}>\sum_{i\in V}k_i^{out}$, \textit{i.e.}~the total number of internal links is larger than half of the number of the external links, since the sum of $k^{in}_i$ is twice the number of internal edges. The similarity with properties 1.) and 2.) of our definition is evident, but instead of comparing absolute numbers, our definition compares absolute numbers to expectation values for these quantities in form of the coefficients of cohesion and adhesion. One of the consequences of Radicchi \textit{et al.'s} definitions is that every union of two communities is also a community. This leads to the strange situation that a community in the ``strong'' or ``weak'' sense can also be an ensemble of disjoint groups of nodes. This paradox may only be resolved if one assumes \textit{a priori} that there exists a hierarchy of communities. The following considerations and examples will show that hierarchies in community structures are possible, but cannot be taken for granted. The representation of community structures by dendograms, therefore, cannot always capture the true community structure. Another definition of communities that implies a hierarchy is that given by Palla \textit{et al.}~ There, a community is interpreted as a set of nodes that can be reached through a clique percolation process. This definition is very strict and focuses more on local structural properties of the graph, whereas the other definitions, including ours, have a link density based interpretation which also makes them more robust to in the case of  ``noisy'' data sets.  

\section{Overlap and Stability of Community Assignments}
\noindent  One cannot generally assume that a community structure of a network is uniquely defined. There may exist several but very different partitions that all have a comparably high value of modularity. Palla \textit{et al.}~\cite{Palla05} have introduced an algorithm to detect overlapping communities by clique percolation and Gfeller \textit{et al.}~have introduced the notion of nodes lying  ``between clusters'' \cite{Gfeller}.  In the framework of this article, the overlap of communities is linked to the degeneracy of the minima of the Hamiltonian. This degeneracy can arise in several ways and we have to differentiate between two different types of overlap: overlap of community structure and overlap of communities. 

\begin{figure}
	\includegraphics[width=8.5cm]{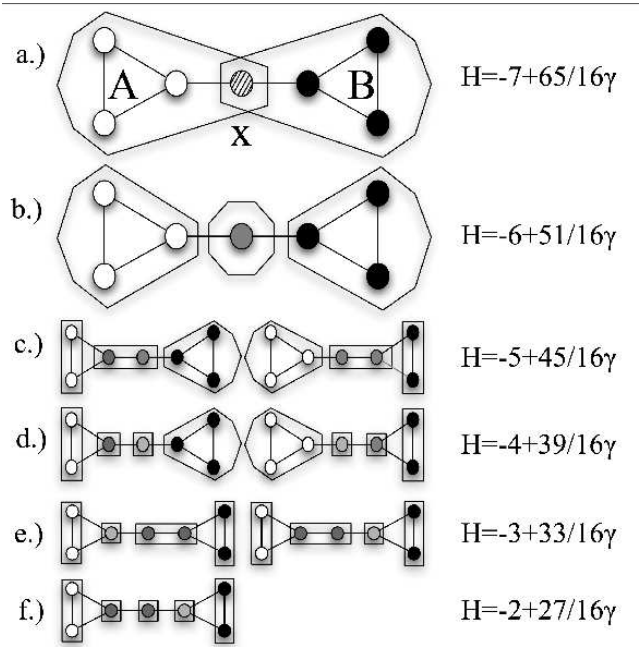}
	\caption{For different values of $\gamma$, different spin configurations minimize the energy to form ground states. For $\gamma<16/63$, the ground state is ferromagnetic. For $16/63<\gamma<8/7$, the two-fold degenerate configuration a.) is the ground state, with node $x$ belonging either to community $A$ or $B$. For $8/7<\gamma<8/3$, configuration b.) shows the non-degenerate ground state. For $\gamma=8/3$, configurations  b.), c.), d.), e.) and f.) all form ground states, but only f.) is ground state for $8/3<\gamma<4$.}
	\label{OverlapExample}
\end{figure}

We have already seen that it is undecidable whether a group of nodes $n_t$ should be member of community $n_s$ or $n_r$, if the coefficients of adhesion are equal for both of these communities. Formally, we find $a_{t,s\backslash t}=a_{tr}$. In this situation, we speak of overlapping communities $n_s$ and $n_r$ with overlap $n_t$, since the number of communities in the network is not affected by this type of degeneracy. Nodes that do not form part of overlaps will always be grouped together and can be seen as the non-overlapping cores of communities. An example of this can be found in Figure \ref{OverlapExample} a.), where communities $A$ and $B$ overlap in node $x$. The ground state at $\gamma=1$ is twofold degenerate with node $x$ belonging either to $A$ or $B$.   

On the other hand, it may be undecidable, if two groups of nodes should be grouped together or apart, if the coefficient of adhesion between them is zero, \textit{i.e.}~there exist as many edges between them as expected from the model $p_{ij}$. Similarly, it may be undecidable, if a group of nodes should form its own community or be divided and the parts joined with different communities, if this can be done without increasing the energy. In these situations, the number of communities in the ground state is not well defined and we cannot speak of overlapping communities, since communities do not share nodes in the degenerate realisations. We will hence refer to such a situation as overlapping community structures. An example of this can be found in Figure \ref{OverlapExample} d.), where the three nodes in groups $A$ and $B$ form either one community as in a.) or two distinct communities of $2$ and $1$ node each. In general, however, both types of overlap may be present in a network.

Since the coefficients of adhesion and cohesion depend on the value of $\gamma$ chosen, one can assess the stability of community structures under the change of this parameter. The network shown in Figure \ref{OverlapExample} illustrates the change of the ground state configuration with $\gamma$.  

\begin{figure*}[t]
	\begin{minipage}[b][5.5cm][t]{5.5cm}
		\fbox{\includegraphics[width=5cm]{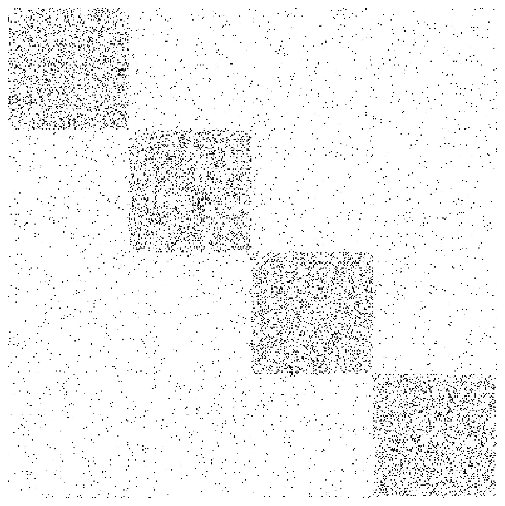}}
	\end{minipage}
	\begin{minipage}[b][5.5cm][t]{5.5cm}
		\fbox{\includegraphics[width=5cm]{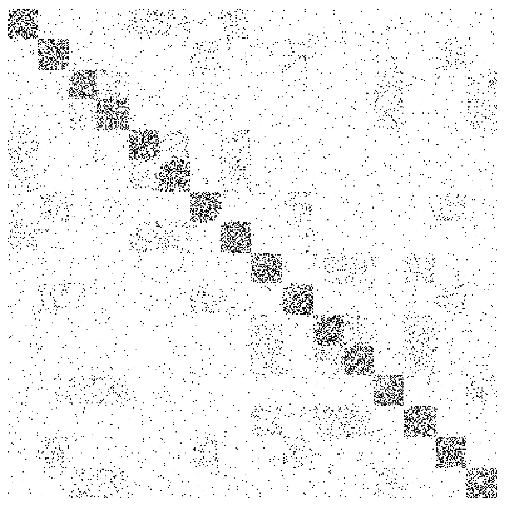}}
	\end{minipage}
	\begin{minipage}[b][5.5cm][t]{5.5cm}
		\fbox{\includegraphics[width=5cm]{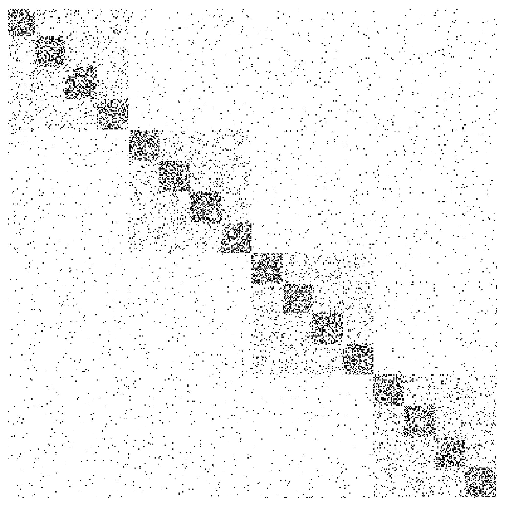}}
	\end{minipage}
\caption{Example of an adjacency matrix for a perfectly hierarchical network. The network consists of four communities, each of which is composed of four sub-communities. Using $\gamma=1$, we find the four main communities (left). With $\gamma=2.2$, we find the 16 sub communities (middle). Link density variations in the off diagonal parts of the adjacency matrix already hint at a hierarchy. The consensus ordering (right) shows, that each of the larger communities is indeed composed of four sub-communities each.}
\label{HierarchieExample}
\end{figure*}

We have already stressed, that properties $1$ through $3$ are also valid for any local minimum of the energy landscape defined by the Hamiltonian and the graph. They only imply that one cannot jump over energy barriers and move into deeper minima using the suggested move set. It may therefore be interesting to study also the local minima and compare them to the ground state. Local minima may be sampled by running greedy optimization algorithms using random initial conditions. This allows for a probabilistic interpretation of the community structure induced by the minima of the Hamiltonian. For correlated energy landscapes, it is known that deeper local minima have larger basins of attraction in the configuration space. The Hamiltonian (\ref{EasyHam}) induces such a correlated energy landscape on the graph, since the total energy is not drastically affected by single spin changes. We therefore expect that the deep local minima will be sampled with higher frequency and that pairs of nodes that are grouped together in deep minima will have larger entries in a co-appearance matrix $C_{ij}$ that keeps track of how frequently node $i$ and $j$ have been grouped together in a local minima for multiple runs of a minimization routine.  A number of examples of co-apperance matrices sampling local energy minima at different values of $\gamma$ have been given in \cite{ReichardtPRL}. 
\begin{figure*}[t]
	\begin{minipage}[b][5.5cm][t]{5.5cm}
		\fbox{\includegraphics[width=5cm]{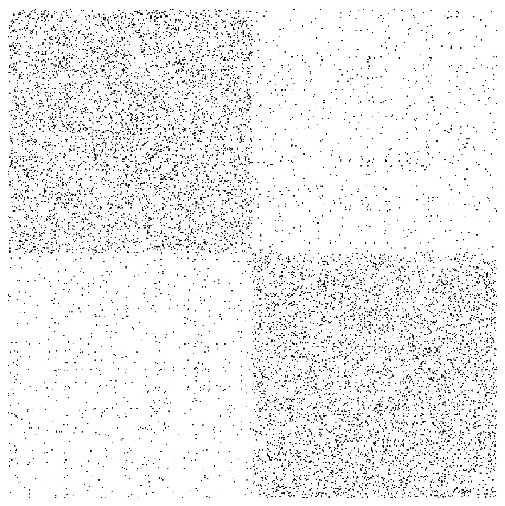}}
	\end{minipage}
	\begin{minipage}[b][5.5cm][t]{5.5cm}
		\fbox{\includegraphics[width=5cm]{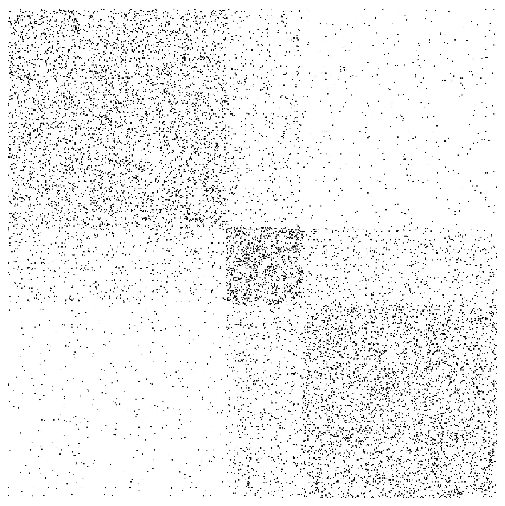}}
	\end{minipage}
	\begin{minipage}[b][5.5cm][t]{5.5cm}
		\fbox{\includegraphics[width=5cm]{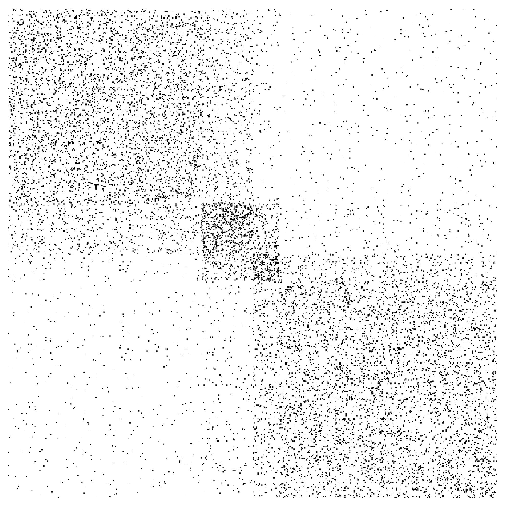}}
	\end{minipage}
\caption{Example of an adjacency matrix for an only partially hierarchical network with overlapping community structure. The network consists of two large communities $A$ and $B$, each of which contains a sub-community $a$ and $b$, which are densely linked with each other, Using $\gamma=0.5$, we find the two large communities (left). With a larger $\gamma=1$, we find the two small sub-communities $a$ and $b$  grouped together.  The consensus ordering (right) shows, that most of the links, that join $A$ and $B$ in fact lie between $a$ and $b$.}
\label{NonHierarchyExample}
\end{figure*}

Here, we shall instead investigate the possible hierarchies of the community structures directly from the adjacency matrix. The ordering of the rows and columns corresponding to nodes of the network is such that between any two nodes that are assigned the same spin state, there never lies a node of different spin. The internal order among the nodes of the same spin state is random. The choice of the ordering of the communities is arbitrary, but some orderings may be more intuitive than others. The link density in the adjacency matrix is directly transformed into grey levels. Since the inner link density of a community is higher than the external, we can distinguish communities as square blocks of darker grey. Different orderings may be combined into a consensus ordering.  That is, starting from a super ordering given, we reorder the nodes within each community according to a second given sub-ordering, \textit{i.e.}~we only change the internal order of the nodes within communities of the super-ordering. 

First, we give an example of a completely hierarchical network. By hierarchical, we mean, that all communities found at a value of $\gamma_2>\gamma_1$ are proper sub-communities of the communities found at $\gamma_1$. In our example, we have constructed a network made of four large communities of $128$ nodes each. Each of these nodes have an average of $7.5$ links to the $127$ other members of their community and $5$ links to the remaining $384$ nodes in the network. Each of these four communities is composed of four sub-communities of $32$ nodes each. Each node has an additional $10$ links to the $31$ other nodes in its sub community. Figure \ref{HierarchieExample} shows the adjacency matrix of this network in different orderings. At $\gamma=1$, the ground state is composed of the four large communities as shown in the left part of Figure \ref{HierarchieExample}. Increasing $\gamma$ above a certain threshold makes assigning different spin states to the 16 sub communities the ground state configuration. The middle part of  Figure \ref{HierarchieExample} shows an ordering obtained with a value of $\gamma=2.2$. We can see, that some of the these sub-communities are more densely connected among each other. Imposing the latter ordering on top of the ordering obtained at $\gamma=1$ then allows to display the full community structure and hierarchy of the network as shown in the right part of Figure \ref{HierarchieExample}. Note that we have \textit{not} used a recursive approach applying the community detection algorithm to separate subgroups. Instead, we have obtained two \textit{independent} orderings which are only compatible with each other, because the network has a hierarchical structure of dense communities composed of denser sub-communities.     

In contrast to this situation, Figure \ref{NonHierarchyExample} shows an example of a network that is only partially hierarchical. The network consists of 2 large communities $A$ and $B$ containing $512$ nodes, which have on average $12$ internal links per node. Within $A$ and $B$, a sub-group of $128$ nodes exists, which we denote by $a$ and $b$, respectively. Every node within this sub-group has $6$ of its $12$ intra-community links with the $127$ other members of this sub-group. The two sub-groups $a$ and $b$ have on average $3$ links per node with each other. Additionally, every node has two links with randomly chosen nodes from the network. From Figure \ref{NonHierarchyExample}, we see that we find the two large communities using $\gamma=0.5$. Maximum modularity, however, is reached at $\gamma=1$ when $a$ and $b$ are joined into a separate community. Only when using the consensus of the ordering obtained at $\gamma=0.5$ and $\gamma=1$, we can understand the full community structure with $a$ and $b$ being subgroups that are responsible for the majority of links between $A$ and $B$. It is understood, that this situation cannot be interpreted as a hierarchy, even though $a$ and $b$ are cohesive subgroups in $A$ and $B$, respectively. We shall now turn to a real world example to see, whether these structural properties can indeed be found outside of artificially constructed examples.

\begin{figure*}
	\begin{tabular}{ll}
	\begin{minipage}[b][10cm][t]{9cm}
			\fbox{\includegraphics[width=8.5cm]{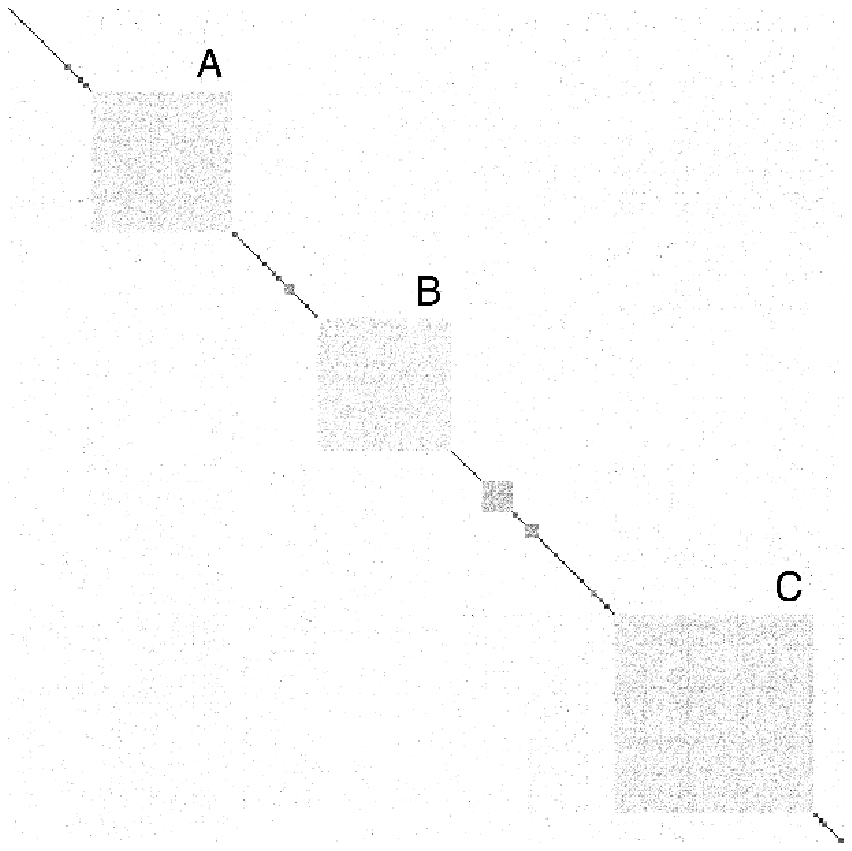}}
			\caption{Adjacency matrix of the co-author network ordered according to the ground state with $\gamma=0.5$.}
			\label{CoAuthorG05}
	\end{minipage} &
	\begin{minipage}[b][10cm][t]{9cm}
			\fbox{\includegraphics[width=8.5cm]{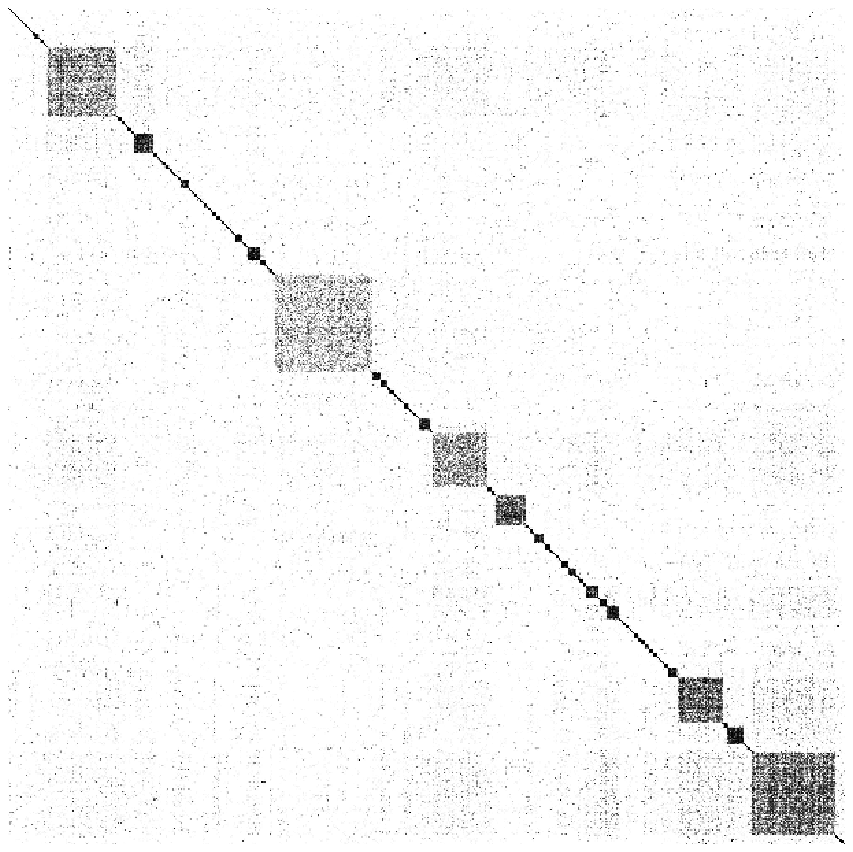}}
			\caption{Adjacency matrix of the co-author network ordered according to the ground state with $\gamma=1$.}
			\label{CoAuthorG1}
	\end{minipage} \\
	\begin{minipage}[b][10.5cm][t]{9cm}
		\fbox{\includegraphics[width=8.5cm]{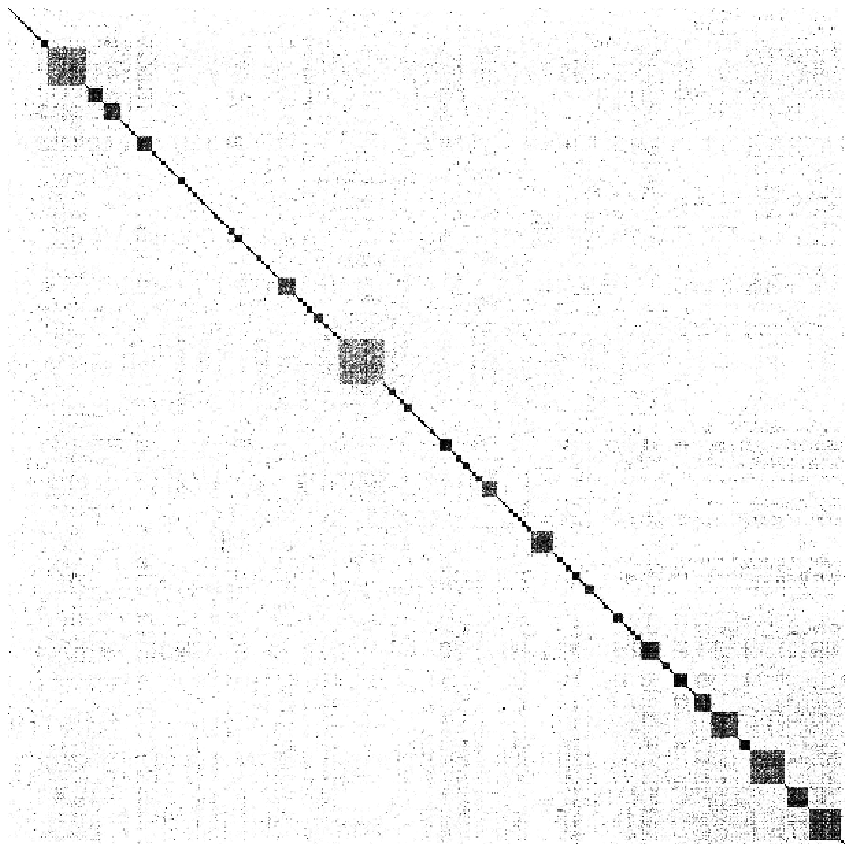}}
			\caption{Adjacency matrix of the co-author network ordered according to the ground state with $\gamma=2$.}
			\label{CoAuthorG2}
	\end{minipage} &
	\begin{minipage}[b][10.5cm][t]{9cm}
			\fbox{\includegraphics[width=8.5cm]{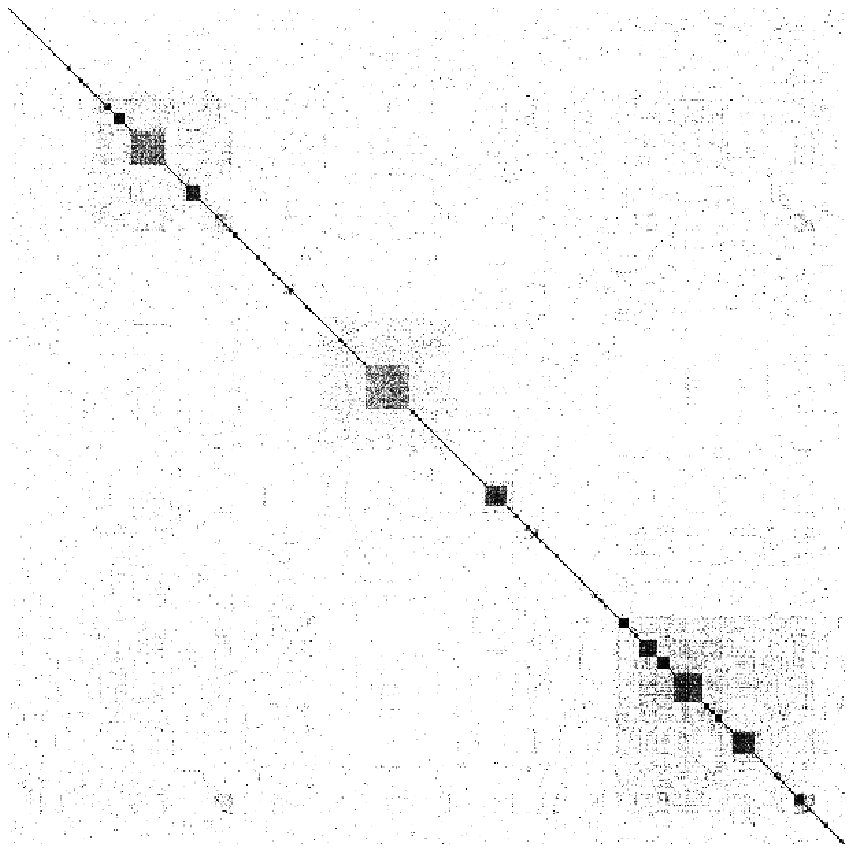}}
			\caption{Adjacency matrix of the co-author network ordered first according to the ground state with $\gamma=0.5$. Within the clusters, the nodes were then ordered again according to the ground state with $\gamma=1$ and within these clusters, the nodes were ordered according to the ground state with $\gamma=2$.\\}
\label{CoAuthorTogether}
	\end{minipage} \\
	\end{tabular}
\end{figure*}

As a real world example, we study the co-authorship network \cite{Warner} of the Los Alamos condensed matter preprint archive, considering articles published between April 1998 and February 2004. This network has also been analyzed by Palla \textit{et al.}~in \cite{Palla05}. Every article induces a complete subgraph between the authors in this network. Since articles with a large number of authors induce very large cliques, every link induced by a single paper of $n$ authors is only given a weight of $1/(n-1)$. After summing the weights for all papers, only links with a weight of $0.1$ and greater were kept, transforming the network into a non-weighted one.    
The network consists of $30,561$ nodes connected by $125,959$ links. There are $668$ connected components, the largest of which has $28,502$ nodes and $123,604$ links. We only work with the largest connected component. The average degree is $\langle k\rangle=8.7$. We then minimize the Hamiltonian (\ref{EasyHam}) using $p_{ij}=k_ik_j/2M$ and $q=500$. Three different values of $\gamma$ were used. For each of the values of $\gamma$, some of the $500$ spin states remained unpopulated, which makes us confident that we provided enough spin states. Figure \ref{CoAuthorG05} shows the adjacency matrix of the co-authorship network with rows and columns ordered according to the ground state at $\gamma=0.5$.  We can distinguish 3 major communities  along the diagonal of the matrix and a large number of smaller communities. Off-diagonal entries in the matrix show, where communities are connected with each other. Figure \ref{CoAuthorG1} shows the same adjacency matrix, but ordered according to the ground state obtained at $\gamma=1$, while Figure \ref{CoAuthorG2} was obtained ordering the adjacendy matrix  according to the ground state obtained at $\gamma=2$. We see, how the increase of $\gamma$ leads to to a higher number of smaller communities and a reduction in size of the major communities as expected.  In Figure \ref{CoAuthorTogether}, we show the adjacency matrix in a consensus ordering of the three single orderings. If the network was hierarchical with respect to $\gamma$, \textit{i.e.}~the communities found for larger values of $\gamma$ are all complete sub-communities of those found at smaller $\gamma$, we should be able to distinguish this from the adjacency matrix in the same manner as in Figure \ref{HierarchieExample}.  From the consensus ordering, we can see that community $A$ from the $\gamma=0.5$ ordering is composed of a number of smaller communities in a somewhat hierarchical manner, while community $B$ seems to consist of a dense core and many adjacent nodes that are gradually removed as $\gamma$ increases. Community $C$ again is decomposed into several smaller subgroups by the consensus ordering that seems to show two levels of hierarchy. The interpretation of the community structure and its hierarchy in terms of research fields is beyond the scope of this article and shall not be attempted here. Rather, we intend to show that both hierarchical and overlapping community structure exists in the link patterns of real world networks and how it can be uncovered.   

\section{Minimizing the Hamiltonian}
\noindent After having studied some properties of the ground state, we now turn to the problem of actually finding it. Though any optimization scheme that can deal with combinatorial optimization problems may be implemented \cite{HartmannOpt, HartmannNewOpt}, we show the use of Simulated Annealing \cite{Kirkpatrick} for this Potts-model, because it yields high quality results, is very general in its application and very simple to program.  The single spin heat bath update rule at temperature $T=1/\beta$ is as follows:
\begin{equation}
p(\sigma_l=\alpha)=\frac{\exp\left(-\beta\mathcal{H}(\{\sigma_{i\neq l},\sigma_l=\alpha\} )\right)}{\sum_{s=1}^{q}\exp\left(-\beta\mathcal{H}(\{\sigma_{i\neq l},\sigma_l=s\})\right)}.
\label{updateSimple}
\end{equation} 
That is, the probability of spin $l$ being in state $\alpha$ is proportional to the exponential of the energy of the entire system with all other spins $i\neq l$ fixed and spin $l$ in state $\alpha$. Since this is costly to evaluate, we pretend that we know the energy of the system with spin $l$ in some arbitrarily chosen spin state $\phi$, which we denote by $\mathcal{H}_{\phi}$. Then we can calculate the energy of the system with $l$ in state $\alpha$ as $\mathcal{H}_{\phi}+\Delta\mathcal{H}(\sigma_l=\phi\to\alpha)$. The energy $\mathcal{H}_{\phi}$ then factors out in (\ref{updateSimple}) and we are left with:
\begin{equation}
p(\sigma_l=\alpha)=\frac{\exp\left\{-\beta\Delta\mathcal{H}(\sigma_{l}=\phi\to\alpha)\right\}}{\sum_{s=1}^{q}\exp\left\{-\beta\Delta\mathcal{H}(\sigma_{l}=\phi\to s)\right\}}.
\label{updateSimple2}
\end{equation} 
The change in energy $\Delta\mathcal{H}(\sigma_{l}=\phi\to\alpha, \phi\neq\alpha)$ is easily calculated for both the models of $p_{ij}$. For the simpler of the two with $p_{ij}=p$, we find:
\begin{widetext}
\begin{eqnarray}
\Delta\mathcal{H}(\sigma_{l}=\phi\to\alpha,\phi\neq\alpha) & = &\sum_{j\neq l}(A_{lj}-\gamma p)\delta(\phi,\sigma_j)-\sum_{j\neq l}(A_{lj}-\gamma p)\delta(\alpha,\sigma_j)\\
& = & \sum_{j\neq l}A_{lj}\delta(\phi,\sigma_j)-\gamma p(n_\phi-1)-\sum_{j\neq l}A_{lj}\delta(\alpha,\sigma_j)+\gamma pn_\alpha\\
& = & a_{l\phi}-a_{l\alpha}.
\label{updateSimple3}
\end{eqnarray}
Here, $n_\phi$ and $n_\alpha$ are the number of nodes in spin state $\phi$ and $\alpha$ respectively, \textit{i.e.}~the size of groups $\phi$ and $\alpha$.  For the model with $p_{ij}=k_ik_j/2M$ we find the following update rule:
\begin{eqnarray}
\Delta\mathcal{H}(\sigma_{l}=\phi\to\alpha,\phi\neq\alpha) & = &\sum_{j\neq l}(A_{lj}-\gamma\frac{k_lk_j}{2M})\delta(\phi,\sigma_j)-\sum_{j\neq l}(A_{lj}-\gamma\frac{k_lk_j}{2M})\delta(\alpha,\sigma_j)\\
& = & \sum_{j\neq l}A_{lj}\delta(\phi,\sigma_j)-\gamma\frac{k_l}{2M}(K_{\phi}-k_l)-\sum_{j\neq l}A_{lj}\delta(\alpha,\sigma_j)+\gamma\frac{k_l}{2M}K_\alpha\\
& = & a_{l\phi}-a_{l\alpha}.
\label{updateSimple4}
\end{eqnarray}
\end{widetext}
Here, again, $K_\phi$ and $K_\alpha$ denote the sum of degrees of nodes in states $\phi$ and $\alpha$, respectively. In both cases, comparing the adhesion of spin $\sigma_l$ with its present community $n_{\phi}$ and all other communities $n_{\alpha}$ the spin state for which the adhesion is largest is assigned the largest probability.  Only local information about the states of the neighbors of a node and some global bookkeeping is necessary. This makes the implementation of a simulated annealing or any other optimization algorithm especially simple and efficient, even though we are dealing with an infinite range spin glass which has non-zero couplings between all pairs of nodes. 

\section{Finding the community around a given node}
\noindent Often, it is desirable not to find all communities in a network, but to find only the community to which a particular node belongs. This may be especially useful if the network is very large and detecting all communities may be time consuming \cite{Wu}. In the framework presented in this article, we can do this using a fast, greedy algorithm. Starting from the node $j$ we are interested in, we successively add nodes with positive adhesion to the group, as long as the adhesion of the community we are forming and the rest of the network decreases. Adding a node $i$ from the rest of the network $r$ to the community $s$ around the start node, the adhesion between $s$ and $r$ changes by:
\begin{equation}
\Delta a_{sr}(i\to s) = a_{ir}-a_{is}.
\end{equation}
For $p_{ij}=p$, this can be written as
\begin{equation}
\Delta a_{sr}(i\to s) = k_{ir}-k_{is} - \gamma p(n_r-1-n_s), 
\end{equation}
where $n_r=N-n_s$ is the number of nodes in the rest of the network, and $n_s$ the number of nodes in the community.
For $p_{ij}=k_ik_j/2M$, the change in adhesion reads:
\begin{equation}
\Delta a_{sr}(i\to s)  =  k_{ir}-k_{is}-\frac{\gamma}{2M}k_i\left(K_r-k_i-K_s\right).\nonumber
\end{equation}
Here, $K_r$ and $K_s$ are the sums of degrees of the rest of the network and the community under study, respectively, and $k_i$ is the degree of node $i$ to be moved from $r$ to $s$, which has $k_{is}$ links connecting it with $s$ and $k_{ir}$ links connecting it with the rest of the network. It is understood that only when the adhesion of $i$ with $s$ is larger than with $r$, the total adhesion of $s$ with $r$ decreases. Equivalent expressions can be found for removing a node $i$ from the community $s$ and rejoining it with $r$.
For $\gamma=1$ and $p_{ij}=k_ik_j/2M$, we have $a_{is}+a_{ir}+2c_{ii}=0$, and $c_{ii}<0$ by definition and close to zero for all practical cases. Then, $a_{is}$ and $a_{ir}$ are either both positive and very small or have opposite sign. Choosing the node that gives the smallest $\Delta a_{rs}$ will then result in adding a node with positive coefficient of adhesion to $s$. It is easy to see, that this ensures a positive coefficient of cohesion in the set of nodes around $j$. 
 
In order to benchmark the performance of this approach, we applied it again to computer generated test networks as done for the algorithm on the entire network in \cite{ReichardtPRL}. We used networks of $128$ nodes, which are grouped into four equal sized communities of size $32$. Each nodes has an average degree of $\langle k\rangle=16$. The average number of links to members of the same community $\langle k_{in}\rangle $ and to members of different communities $\langle k_{out}\rangle $ is then varied, but always ensuring   $\langle k_{in}\rangle+\langle k_{out}\rangle=\langle k\rangle$. Hence, decreasing $k_{in}$ renders the problem of community detection more difficult. Starting from a particular node, we are interested in the performance of the algorithm in discovering the community around it. We measure the percentage of nodes that are correctly identified as belonging to the community around the start node as sensitivity and the percentage of nodes that are correctly identified as \textit{not} belonging to the community as specificity.

\begin{figure}[t]
	\includegraphics[width=8cm]{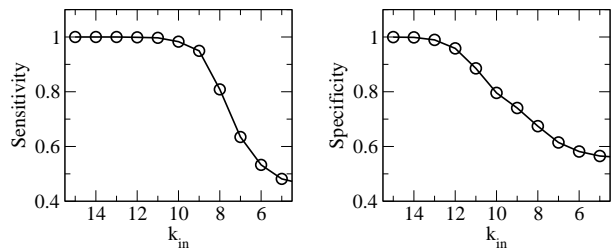}
\caption{Benchmark of the algorithm for discovering the community around a given node in networks with known community structure. We used networks of $128$ nodes and four communities. The average degree of the nodes was fixed to $16$, while the average number of intra-community links $\langle k_{in}\rangle $ was varied. Sensitivity measures the fraction of nodes correctly assigned to the community around the start node, while specificity measures the fraction of nodes correctly kept out of the community around the start node.}
\label{SingleNodeBM}
\end{figure}

Figure \ref{SingleNodeBM} shows the results obtained for different values of $\langle k_{in}\rangle$ at $\gamma=1$ and using $p_{ij}=k_ik_j/2M$ as model of the connection probability. We note, that this approach performs rather well for a large range of $\langle k_{in}\rangle$ with good sensitivity and specificity. In contrast to the benchmarks for running the simulated annealing on the entire network as shown in \cite{ReichardtPRL}, we obtain a sensitivity that is generally larger than the specificity. This shows, that running the simulated annealing on the entire network tends to mistakenly group things apart, that do not belong apart by design, while constructing the community around a given node, tends to group things together that do not belong together by design. This behavior is understandable, since working on the entire network amounts to effectively implementing a divisive method, while starting from a single node means implementing an agglomerative method.   

\section{Expectation Values for the Modularity}
\begin{figure}[h] 
\begin{center}
\includegraphics[width=8cm]{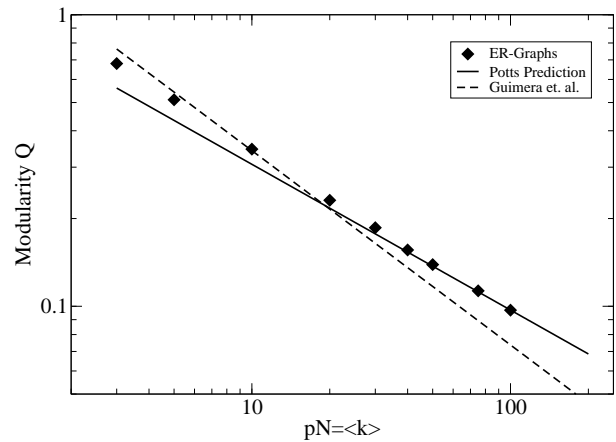}
\end{center}
\caption{Modularity of Erd\H{o}s R\'{e}nyi random graphs with average connectivity $pN=\langle k\rangle$ compared with the estimation from equation (\ref{ModularityEstimation}). For the experiment, random graphs with $N=10000$ were used.}
\label{TheoPred}
\end{figure}

\noindent In order to assess the statistical significance of the modularities found with any algorithm, it is necessary, to compare them with expectation values for random networks. This is of course always possible by rewiring the network randomly \cite{Maslov02}, keeping the degree distribution invariant and then running a community detection algorithm again, comparing the result to the original network. This method, however, can only give an answer to what a particular community detection algorithm may find in a random network and hence depends on the very method of community detection used. Much better seems a method to compare the results of a community detection algorithm with a theoretical result, obtained independently of any algorithm. 
We have already seen, that the problem of community detection can be mapped onto finding the ground state of an infinite range spin glass. In the limit of large $N$, the local field distribution of infinite range spin glasses is Gaussian and can hence be characterized by only the first two moments of the coupling distribution, the mean and the variance. The couplings used in the study of modularity are $J_{ij}=A_{ij}-\gamma p_{ij}$ which have a mean independent of the particular form of $p_{ij}$:
\begin{equation}
J_0=(1-\gamma)p
\end{equation}
which is zero in the case of the ``natural partition'' at $\gamma=1$. The variance amounts to:
\begin{equation}
J^2=p-(2\gamma-\gamma^2)\langle p^2\rangle.
\end{equation}
Since the mean of the coupling distribution couples to the magnetization of the ground state, all coupling distributions with zero mean will have zero magnetization in the ground state. Hence, for a random graph we expect maximum modularity for an equi-partition. A number of well known results exist in the literature for equi-partitions. Fu and Anderson \cite{FuAnd} have given results for bi-partitionings and Kanter and Sompolinsky for q-partitionings \cite{KanterSompQPart}. With these, we can write immediately for the modularity at $\gamma=1$: 
\begin{equation}
Q=-\frac{1}{M}\mathcal{H}_{GS}=\frac{N^{3/2}}{M}J\frac{U(q)}{q},
\label{PottsModularity}
\end{equation}
where $U(q)$ is the ground state energy of a q-state Potts model with Gausssian couplings of zero mean and variance $J^2$. For large $q$, we can approximate $U(q)=\sqrt{q\ln q}$. In Table \ref{PottsGS} we give some small values of $q$ obtained by using the exact formula for calculating $U(q)$ from  \cite{KanterSompQPart}.
\begin{table}[h]
\begin{tabular}{|l||c|c|c|c|c|c|c|c|}
\hline
$q$ & $2$ & $3$  & $4$ & $5$ & $6$ & $7$ & $8$ & $9$\\
$U(q)/q$ & $0.384$ & $0.464$ & $0.484$ & $0.485$ & $0.479$ & $0.471$ & $0.461$ & $0.452$\\
\hline
\end{tabular}
\caption{Values of $U(q)/q$ for various values of $q$ obtained from  \cite{KanterSompQPart}, which can be used to approximate the expected modularity with equation (\ref{PottsModularity}).}
\label{PottsGS}
\end{table}
We see, that maximum modularity is obtained at $q=5$, though the value of $U(q)/q$ for $q=4$ is not much different from it. This qualitative behavior, that dense random graphs tend to cluster into only a few large communities is confirmed by our numerical experiments. 
By rewriting $M=pN^2/2$ and under the assumption of $p_{ij}=p$ as in the case of  Erd\H{o}s R\'{e}nyi (ER) random graphs \cite{Erdos}, we can further simplify equation (\ref{PottsModularity}) and write for the maximum value of the modularity of a ER random graph with connection probability $p$ and $N$ nodes:
\begin{equation}
Q=0.97\sqrt{\frac{1-p}{pN}}
\label{ModularityEstimation}
\end{equation}
where we have already made use of the fact, that $q=5$ makes the modularity maximal. Figure \ref{TheoPred} shows the comparison of equation (\ref{ModularityEstimation}) and experiments where we have numerically maximized the modularity using a simulated annealing approach as described in an earlier section. 
We see, that the prediction fits the data well for dense graphs and that modularity decays as a function of $(pN)^{-1/2}$ instead of $(2/pN)^{2/3}$ as proposed in \cite{Guimera}. 

While the value of $Q$ for random graphs from the Potts spin glass is rather close to the actual situation for sparse random graphs, the number of communities, at which maximum modularity is achieved is not. In \cite{Guimera}, it had already been shown, that the number of communities for which the modularity reaches a maximum is $\sqrt{N}$ for treelike networks with $\langle k\rangle=2$. Unfortunately, no plot was given for the number of communities found in denser networks. Our numerical experiments on large Erd\H{o}s R\'{e}nyi random graphs also show, that the number of communities found in sparse networks tends to increase as $\langle k\rangle$ decreases. 

Even though we have seen that in general, recursive bi-partitioning will not lead to an optimal community assignment, we shall still use this approach for random graphs. Maximum modularity for random graphs is achieved for equipartitions. Partitioning the network recursively until no further improvement of $Q$ is possible allows us to find the number of communities in a random graph. The number of cut edges $\mathcal{C}=\mathcal{C}(N,M)$ in any partition, will be a function of the number of nodes in the remaining part and the number of connections within this remaining part and their distribution. We note, that the $M$ connections will be distributed into internal and external links per node $k_{in}+k_{out}=k$. This allows us to write $\mathcal{C}=N\langle k_{out}\rangle/2$ for a bi-partition. After each partition, the number of internal connections a node has decreases due to the cut. We use these results in order to approximate the number of cut edges after $b$ recursive bi-partitions which lead to $2^b$ parts:
\begin{equation}
\mathcal{C}=\sum_{t=1}^{b}2^{t-1}\frac{N}{2^t}\langle k_{out,t}\rangle=\sum_{t=1}^b \frac{N}{2}\langle k_{out,t}\rangle
\label{RECURSIVE}
\end{equation}
where $\langle k_{out,t}\rangle $ is the average number of external edges a node gains after cut $t$. Since for an Ising-model, the ground state energy is $-E_{GS}=M-2C$ we find:
\begin{equation}
\langle k_{in}\rangle=\frac{\langle k\rangle}{2}-E_{GS}(\langle k\rangle)=\langle k\rangle-\langle k_{out}\rangle.
\label{KIN}
\end{equation}
This shows, that for any bi-partition, we can, on average, always satisfy more than half of the links of every node on average. This means also, that any bi-partition will satisfy the definition of community given by Radicci \cite{Radicchi} at least on average, which further means, that every random graph has - at least on average - a community structure, assuming Radicci's definition of community in a strong sense $(k_{in}>k_{out})$ for every node of the random graph. The definition of community in a weak sense $\sum_i k_i^{in}>\sum_i k_i^{out}$ can always be fulfilled in a random graph. 

\begin{figure}[t] 
\begin{center}
\includegraphics[width=8cm]{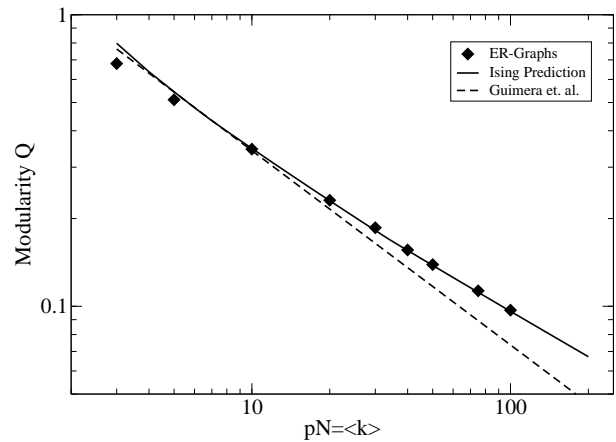}
\end{center}
\caption{Modularity of Erd\H{o}s R\'{e}nyi random graphs with average connectivity $pN=\langle k\rangle$ compared with the estimation from equation \ref{IsingPrediction}. For the experiment, random graphs with $N=10000$ were used.}
\label{IsingPred}
\end{figure}

From (\ref{KIN}) we can then calculate the total number of edges cut after $t$ recursions according to (\ref{RECURSIVE}) using results of Fu and Anderson \cite{FuAnd} again who find for a bi-partition:
\begin{equation}
\mathcal{C}=\frac{M}{2}\left[1-c\sqrt{\frac{1-p}{pN}}\right].
\label{FuAndCut}
\end{equation}  
with a constant of $c=1.5266\pm 0.0002$. We can write
\begin{equation}
\langle k_{in}\rangle=\frac{pN+c\sqrt{pN(1-p)}}{2}=pN-\langle k_{out}\rangle
\end{equation}
from which we can calculate (\ref{RECURSIVE}) substituting $pN$ with the appropriate $\langle k_{in}\rangle$ in every step of the recursion. The modularity can then be written:
\begin{equation}
Q=\frac{2^b-1}{2^b}-\frac{1}{\langle k\rangle}\sum_{t=1}^b\langle k_{out,t}\rangle.
\label{IsingPrediction}
\end{equation}
Now we only need to find the number of recursions $b$ that maximizes $Q$. Since the optimal number of recursions will depend on $pN$, we also find an estimation of the number of communities in the network. 
Figure \ref{IsingPred} shows a comparison between the theoretical prediction of the maximum modularity that can be obtained from equation (\ref{IsingPrediction}).  The improvement of (\ref{IsingPrediction}) over (\ref{ModularityEstimation}) must be due to the possibility of having larger numbers of communities, since (\ref{FuAndCut}) also assumes a Gaussian distribution of local fields, which is a rather poor approximation for  the sparse graphs under study. Again, we find that the modularity behaves asymptotically like $k^{-1/2}$ as already predicted from the Potts spin glass and contrary to the estimation in \cite{Guimera}.  

\begin{figure}[t] 
\begin{center}
\includegraphics[width=8cm]{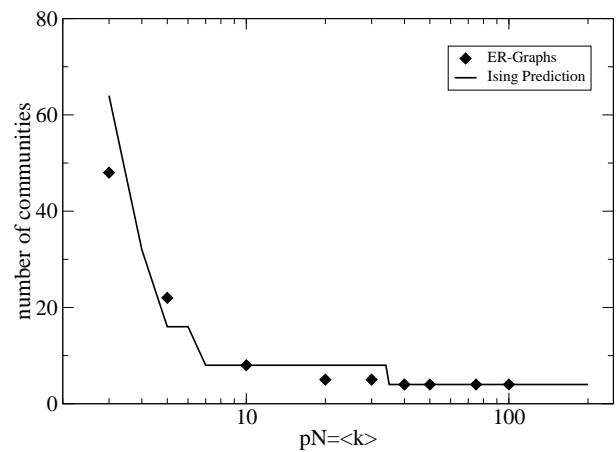}
\end{center}
\caption{Number of communities found in Erd\H{o}s R\'{e}nyi random graphs with average connectivity $pN=\langle k\rangle$ compared with the estimation from equation \ref{IsingPrediction}. For the experiment, random graphs with $N=10000$ were used.}
\label{ModulePred}
\end{figure}

Figure \ref{ModulePred} shows the comparison of the number of communities estimated from (\ref{IsingPrediction}) and the numerical experiments on random graphs. The good agreement between experiment and prediction is interesting, given the fact, that (\ref{IsingPrediction}) allows only powers of two as the number of communities. For dense graphs, the Potts limit of only a few communities is recovered. We see, that sparse random graphs cluster into a large number of communities, while dense random graphs cluster into only a hand full of large communities. Most importantly, sparse random graphs exhibit very large values of modularity. These large values are only due to their sparseness and \textit{not} due to small size. We also stress that statistically significant modularity must exceed the expectation values of modularity obtained  from a suitable null model of the graph. If this null model is an Erd\H{o}s R\'{e}nyi random graph, then there is very little improvement possible over the values of modularity obtained for the null model for sparse graphs.   
     

\section{Conclusion}
\noindent In this article, we have tried to elucidate some of the general properties of the problem of community detection in complex networks. We have shown, that it can be mapped onto finding the ground state of an infinite range Potts spin glass from a very simple and general one parameter \textit{ansatz}, which is also valid for weighted networks and directed networks. We could show that our \textit{ansatz} leads to known modularity measures in a natural way. We have introduced the concept of cohesion and adhesion into the terminology of networks as a measure of the degree to which groups of nodes belong together or apart in a community structure. From the properties of the ground state as the minimal energy or maximally modular configuration, we could deduce a number of properties that define a community. By studying the ground state structure and its changes under parameter variation, we could also show, how hierarchical and overlapping community structures manifest themselves. Comparisons of our with other definitions of communities were given. We have provided efficient update rules for single spin heat bath simulated annealing algorithms that allow to optimize the spin configuration of an infinite range system by using solely sparse local information and some global bookkeeping. We have extended the algorithm of finding the entire community structure of the whole network to finding only the community around a given node and we have given benchmarks for the performance of this extension. Finally, we have summarized known results from the theory of infinite range spin glasses in order to shed some light on the problem of community detection in Erd\H{o}s R\'{e}nyi random graphs. We have seen, that sparse ER random graphs may show very large modularities and that the expected modularity of an ER random graph decays as $\sqrt{1/\langle k\rangle}$ independent of the size of the graph.  Further, we have seen, that sparse ER random graphs tend to cluster into many small communities, while for dense random graphs, maximum modularity is achieved for a very small number of communities only, which is independent of the average degree of the network. We stress the importance of comparing the values of modularity found in real world networks with expectation values of appropriate null models in order to assess their statistical significance. Only graphs which lead to modularities larger than the expectation value should be called modular. In this respect, it is understood that Erd\H{o}s R\'{e}nyi random graphs contain communities, but this alone does not make these graphs modular. 

\section{Acknowledgements}
The authors would like to thank Stefan Braunewell for a careful reading of the manuscript as well as Michele Leone, Ionas Erb and Andreas Engel for many helpful hints and discussions. Also, we would like to thank G. Palla for letting us use the co-author data. 
\bibliography{../../BibTex_Citations}

\end{document}